\documentclass[12pt]{article}
\usepackage{graphicx}
\usepackage{amsmath}
\usepackage{amssymb}
\usepackage{enumitem}
\usepackage{latexsym}
\begin{document}
\vskip 2cm

\def\warning#1{\begin{center}
\framebox{\parbox{0.8\columnwidth}{\large\bf #1}}
\end{center}}

\begin{center}

{\large {\bf Augmented Superfield Approach to Gauge-invariant\\ Massive 2-Form Theory}}

\vskip 2.5 cm

{\sf{ \bf R. Kumar$^1$ and S. Krishna$^2$}}\\
\vskip .1cm
{\it $^1$Department of Physics \& Astrophysics,\\ University of Delhi, New Delhi--110007, India}\\
\vskip .1cm
{\it $^2$Indian Institute of Science Education and Research Mohali, \\
 Sector 81,  SAS Nagar,  Manauli, Punjab--140306,  India}\\
\vskip .2cm
{\tt{E-mails: raviphynuc@gmail.com; skrishna.bhu@gmail.com}}\\
\end{center}

\vskip 2.5 cm

\noindent

\noindent
{\bf Abstract:}
We discuss the complete sets of the off-shell nilpotent (i.e. $s^2_{(a)b} = 0$) and absolutely anticommuting 
(i.e. $s_b\, s_{ab} + s_{ab} \, s_b = 0$) Becchi--Rouet--Stora--Tyutin (BRST) $(s_b)$ and anti-BRST $(s_{ab})$ 
symmetries for the $(3+1)$-dimensional $(4D)$ gauge-invariant massive 2-form theory within the framework 
of augmented superfield approach to BRST formalism. In this formalism, we obtain the coupled (but equivalent) 
Lagrangian densities which respect both BRST and anti-BRST symmetries on the constrained hypersurface defined  by the 
Curci--Ferrari type conditions. The absolute anticommutativity property of the (anti-) BRST transformations (and corresponding generators) 
is ensured by the existence of the Curci--Ferrari type conditions which emerge very naturally in this formalism. Furthermore, the gauge--invariant 
restriction plays a decisive role in deriving the {\it proper} (anti-)BRST transformations for the St{\"u}ckelberg-like vector field. 


\newpage


\section{Introduction}


The antisymmetric 2-form $B^{(2)} = \frac{1}{2!}\, (dx^\mu \wedge dx^\nu)\,B_{\mu\nu}$ gauge field $B_{\mu\nu} (= - B_{\nu\mu})$ \cite{ogi,kalb} 
has paved a great deal 
of interest of the theoretical physicists during past few decades  because of its relevance in the realm of (super-)string theories \cite{pol1,pol2}, 
(super-)gravity theories \cite{sal}, dual description of a massless scalar field \cite{des,aur} and modern 
developments in noncommutative geometry \cite{sei}. It has also been quite popular in the mass generation of the 1-form  
$A^{(1)} = dx^\mu A_\mu$ gauge field $A_\mu$, without taking 
any help of Higgs mechanism, where 2-form and 1-form gauge fields merged together in a particular fashion through a well-known topological 
$(B \wedge F)$ term \cite{crem,ami1,ami2,dsh,pio,ami3}.

The Becchi--Rouet--Stora--Tyutin (BRST) formalism is one of the most elegant and intuitively appealing theoretical approaches to covariantly quantizing
gauge theories  \cite{brs1,brs2,brs3,brs4}. The gauge symmetry is always generated by the first-class constraints present in a 
given theory, in Dirac's terminology \cite{dira,sund}. In the BRST formalism, the classical local gauge symmetry of a given physical theory is traded  
with two global BRST and anti-BRST symmetries at the quantum level \cite{cf,oji}. These symmetries obey two key properties: (i) nilpotency of 
order two, and (ii) absolute anticommutativity. The first property implies that these symmetries are fermionic in 
nature whereas second property shows that they are linearly independent of each other.
In the literature, it has been shown that only the BRST symmetry is not sufficient to yield the ghost 
decoupling from the physical subspace of the total quantum Hilbert 
space of states. The addition of nilpotent anti-BRST symmetry plays an important role in removing the unphysical ghost degeneracy \cite {hwa}. 
Thus, the anti-BRST symmetry is not just a decorative part; rather, it
is an integral part of this formalism and  plays a fundamental role in providing us with the consistent BRST quantization.

The superfield approach to BRST formalism is the theoretical approach that provides the geometrical origin as well as deep 
understanding about the (anti-)BRST symmetry transformations \cite{bt1,bt2,del}. The Curci--Ferrari condition \cite{cf}, 
which is a hallmark of the non-Abelian 1-form gauge theory, emerges very naturally as an off-shoot of the superfield formalism.  
This condition plays a central role in providing the absolute
anticommutativity  property of the (anti-)BRST transformations and also responsible for the derivation of the coupled (but equivalent) Lagrangian densities. 
In recent years, the 
``augmented" superfield formalism,  an extended version of Bonora--Tonin superfield formalism, has been applied to the interacting gauge systems 
such as (non-)Abelian 1-form gauge theories interacting with Dirac fields \cite{rpm1,rpm2,rpm3,rpm4,rpm5} and complex scalar fields \cite{rpm6,rpm7}, gauge-invariant version of the self-dual chiral boson \cite{rpm8},
4D Freedman--Townsend model \cite{rpm9}, 3D Jackiw--Pi model \cite{gk1}, vector Schwinger model in 2D \cite{gk2} and modified version of 2D Proca theory  
\cite{sam}. In this approach, the celebrated horizontality condition and gauge-invariant restrictions are blend together in a physically meaningful manner to derive the proper off-shell nilpotent and absolutely anticommuting  (anti-)BRST symmetry transformations.

As far as the quantization of the 4D (non-)Abelian 2-form gauge theories is concerned, the canonical and BRST quantizations have been
carried out \cite{tak,kau,ft,mb,biz,ami4}. The 2-form gauge theory is a reducible theory and, thus, requires ghost for ghost in the 
latter quantization scheme. In the non-Abelian case, a compensating auxiliary vector field is required for the consistent quantization 
as well as in order to avoid the well-known no-go theorem \cite{hen}. In fact, this auxiliary field is needed to close the symmetry 
algebra and, thus, the theory respects the vector gauge symmetry present in the theory. Furthermore, within the framework of BRST formalism, 
the free Abelian 2-form gauge theory in $(3 +1)$-dimensions of spacetime provides a field-theoretic  model for the  Hodge theory where all 
the de Rham cohomological operators $(d, \, \delta,\, \Delta)$ and Hodge duality ($*$) operation of differential geometry find their physical 
realizations  in the language of  the continuous symmetries and discrete symmetry, respectively \cite{hms,gm}. In addition, it has also 
been shown that the free Abelian 2-form gauge theory,  within the framework of BRST formalism, provides a new kind of  quasi-topological 
field theory (q-TFT) which captures some features of Witten type TFT and a few aspects of Schwarz type TFT \cite{rpm10}.

We have also studied the 4D topologically massive (non)-Abelian 2-form theories where 1-form gauge bosons acquire 
mass through a topological $(B \wedge F)$ term without spoiling the gauge invariance of the theory. With the help of superfield formalism, we
have derived the off-shell nilpotent as well as absolutely  anticommuting (anti-)BRST transformations  and also shown that the topological 
$(B \wedge F)$ term remains unaffected by the presence of the Grassmannian variables when we generalize it on the ($4, 2$)-dimensional 
supermanifold \cite{gkr,kam}.   
In the non-Abelian case, we have found some novel observations. For the sake of brevity, the conserved and nilpotent 
(anti-)BRST charges do not generate the proper (anti-)BRST transformations for the compensating auxiliary vector field \cite{rk1,rk2}. Moreover, 
in contrast to the Nakanishi--Lautrup fields, the nilpotency and absolute anticommutativity properties of the (anti-)BRST 
transformations also fail to produce the correct (anti-)BRST symmetry transformations for the compensating auxiliary field.

The contents of our present investigation are organized as follows. In Sect. 2, we briefly discuss about the 4D massive 2-form 
theory and its constraints structure. Section 3 is devoted to the coupled (but equivalent) Lagrangian densities that respect 
the off-shell nilpotent (anti-)BRST symmetries. We discuss the salient features of the Curci--Ferrari type conditions in this section, too. 
In Sect. 4, we discuss the 
conserved charges as the generator of the off-shell nilpotent (anti-)BRST transformations. The global continuous ghost-scale symmetry 
and BRST algebra among the symmetry transformations (and corresponding generators) are shown in Sect. 5. 
Section 6 deals with the derivation of the proper (anti-)BRST symmetry transformations with the help of augmented superfield formalism.    
We capture the (anti-)BRST invariance of the coupled Lagrangian densities in terms of the superfields and Grassmannian translational generators
in Sect. 7. Finally, in Sect. 8, we provide the concluding remarks.

In Appendix A, we show an explicit proof of the anticommutativity of the conserved (anti-)BRST charges.


\section{Preliminaries: $(3+1)$-dimensional massive Abelian 2-form theory} 

We begin with the $(3 + 1)$-dimensional  (4D)  massive Abelian 2-form theory which is 
described by the following Lagrangian density\footnote{We adopt the conventions and notations such that the 4D flat Minkowski metric endowed with 
mostly negative signatures: $\eta_{\mu\nu} = \eta^{\mu\nu} = \text{diga}\,(+1, -1, -1, -1)$. Here, the Greek indices $\mu, \nu, \kappa,... = 0, 1, 2, 3$ 
correspond to the spacetime directions, whereas the Latin  indices $i, j, k,... = 1,2,3$  stand for the space directions only. We also follow the convention:
$\frac{\delta B_{\mu\nu}}{\delta B_{\kappa \sigma}} = \frac{1}{2!}\,(\delta^\kappa_\mu\, \delta^\sigma_\nu - \delta^\kappa_\nu \, \delta^\sigma_\mu).$}   
\begin{eqnarray}
{\cal L} = \frac{1}{12}\, H^{\mu\nu\eta}\, H_{\mu\nu\eta}  - \frac{m^2}{4}\, B^{\mu\nu}\, B_{\mu\nu},
\end{eqnarray}
where the totally antisymmetric 3-form $H^{(3)} = \frac{1}{3!}\,(dx^\mu \wedge dx^\nu \wedge dx^\eta) H_{\mu\nu\eta}$ defines
the curvature tensor $H_{\mu\nu\eta} = \partial_\mu B_{\nu\eta} 
+ \partial_\nu B_{\eta\mu}+ \partial_\eta B_{\mu\nu}$  for the Abelian 2-form $B^{(2)} = \frac{1}{2!}\,(dx^\mu \wedge dx^\nu) B_{\mu\nu}$
antisymmetric field $B_{\mu\nu}$. The 3-form curvature  $H^{(3)} = d B^{(2)}$ 
owes its origin in the exterior derivative $d = dx^\mu\,\partial_\mu$ (with $d^2 = 0$). In the above, $m$ represents a constant mass parameter.

It is evident that due to the existence of mass term, the Lagrangian density does not respects the  following gauge symmetry:
\begin{eqnarray}
&& \delta B_{\mu\nu} = \partial_\mu \Lambda_\nu(x) - \partial_\nu \Lambda_\mu(x),
\end{eqnarray} 
where $\Lambda_\mu (x)$ is an infinitesimal local vector gauge parameter. In fact, the above Lagrangian density transforms as 
$\delta {\cal L} = - m^2\, B^{\mu\nu}\, (\partial_\mu \Lambda_\nu)$. The basic reason behind this observation is that the above
Lagrangian density is endowed with the second-class constraints, in language of Dirac's prescription for the classification 
scheme of constraints \cite{dira,sund},  namely;
\begin{eqnarray}
&& \chi^i =  \Pi^{0i} \approx 0, \qquad \xi^i =  - (2\partial_j \Pi^{ij} + m^2\, B^{0i}) \approx 0,
\end{eqnarray}
where $\Pi^{0i}$ and $\Pi^{ij}$ are the canonical conjugate momenta corresponding to the dynamical fields $B_{0i}$ and $B_{ij}$, respectively.
Here, the symbol `$\approx$' defines weak equality in the sense of Dirac. Due to the existence of mass term in 
the Lagrangian density, both constraints belong to the category of second-class constraints as one can check that 
the primary ($\chi^i$) and secondary  ($\xi_j$) constraints lead to a non-vanishing Poisson bracket:  
$\big[\chi^i(\vec x, t), \, \xi_j (\vec x', t)\big] = m^2 \,\delta^i_j\, \delta^3 (\vec x - \vec x')$. 
Thus, the mass term in the Lagrangian density 
spoils the gauge invariance. However, on one hand, the gauge invariance can be restored by setting mass parameter equal to zero (i.e. $m = 0$).
But this leads to the massless 2-form gauge theory.   
On other hand, we can restore the gauge invariance by exploiting the power and strength of the well-known St{\"u}ckelberg technique 
(see, e.g. \cite{stu,rue} for details).  
Thus, we re-define the field $B_{\mu\nu}$ as
\begin{eqnarray}
&& B_{\mu\nu} \longrightarrow B_{\mu\nu} = B_{\mu\nu} - \frac{1}{m}\, \Phi_{\mu\nu},
\end{eqnarray} 
where $\Phi_{\mu\nu} = (\partial_\mu \phi_\nu - \partial_\nu \phi_\mu)$ and $\phi_\mu$ is the St{\"u}ckelberg-like vector 
field. As a consequence,  we obtain the following gauge-invariant St{\"u}ckelberg-like Lagrangian density for the massive 
2-form theory \cite{hari1,hari2}:
\begin{eqnarray}
{\cal L}_s &=& \frac{1}{12} \, H^{\mu\nu\eta}H_{\mu\nu\eta}
- \frac{m^2}{4} \,B^{\mu\nu}B_{\mu\nu} - \frac{1}{4}\, \Phi^{\mu\nu}\Phi_{\mu\nu} +\frac{m}{2}\, B^{\mu\nu}\Phi_{\mu\nu}. 
\end{eqnarray}
Here $\Phi_{\mu\nu}$ defines the curvature for the St{\"u}ckelberg-like vector field $\phi_\mu$. In the language of differential 
form, we can write $\Phi^{(2)} = d \phi^{(1)} = \frac{1}{2!} \,(dx^\mu \wedge dx^\nu)\, \Phi_{\mu\nu}$. We, interestingly,  point out that the 
above Lagrangian density and the Lagrangian density for the 4D topologically massive $(B\wedge F)$ theory have shown to be 
equivalent by Buscher's duality procedure  \cite{hari1,hari2}. Furthermore, due to the introduction of St{\"u}ckelberg-like vector field, 
the second-class constraints get converted into the first-class constraints \cite{dira,sund}. These 
first-class constraints are listed as follows:
\begin{eqnarray}
&&\Theta = \Pi^0 \approx 0, \;\qquad  \qquad \Theta^i =\Pi^{0i} \approx 0, \nonumber\\
&& \Sigma = \partial_i \Pi^i \approx 0, \; \qquad\quad\;\; \Sigma^i = - \big(2\,\partial_j \Pi^{ij} + m \, \Pi^i \big) \approx 0,
\end{eqnarray} 
where $\Pi^0$ and $\Pi^i$ are canonical conjugate momenta corresponding to the fields $\phi_0$ and $\phi_i$, respectively. It is elementary to check that 
the Poisson brackets among all the first-class constraints turn out to be zero. Further, the first-class constraints 
$\Sigma$ and $\Sigma^i$ are not linearly independent. They are related as $\partial_i \Sigma^i + m \,\Sigma = 0$ which implies that the Lagrangian (5)
describes a reducible gauge theory \cite{hari2}. 
These first-class constraints are the generators of  two independent local and continuous gauge symmetry transformations, namely;
\begin{eqnarray}
&& \delta_1 \phi_\mu = \partial_\mu \Omega, \qquad\qquad\qquad \qquad \delta_1 B_{\mu\nu} =0,\nonumber\\
&& \delta_2 B_{\mu\nu} = - \big(\partial_\mu \Lambda_\nu - \partial_\nu \Lambda_\mu \big), \qquad \delta_2 \phi_\mu = - m \Lambda_\mu,
\end{eqnarray}
where the Lorentz scalar  $\Omega(x)$ and Lorentz vector $\Lambda_\mu(x)$  are the local gauge parameters. 
It is straightforward to check that under above the gauge transformations, the Lagrangian density remains invariant
 (i.e. $\delta_1 \,{\cal L}_s = 0$ and  $\delta_2 \,{\cal L}_s = 0$).
As a consequence, the combined gauge symmetry transformations $\delta = (\delta_1 + \delta_2)$ also leave the Lagrangian density $({\cal L}_s)$ invariant.


\section{Coupled Lagrangian densities: off-shell nilpotent and absolutely anticommuting (anti-)BRST symmetries}

The coupled (but equivalent) Lagrangian densities for the 4D St{\"u}kelberg-like massive Abelian 2-form  theory
incorporate the gauge-fixing and Faddeev--Popov ghost terms within the framework of BRST formalism. 
In full blaze of glory, these Lagrangian densities (in the Feynman gauge) are given as follows:  
\begin{eqnarray}
{\cal L}_B &=& \frac{1}{12} \, H_{\mu\nu\eta} H^{\mu\nu\eta}
-\frac{1}{4}\,m^2\, B_{\mu\nu} B^{\mu\nu} - \frac{1}{4}\, \Phi_{\mu\nu} \Phi^{\mu\nu} +\frac{1}{2}\, m\, B_{\mu\nu} \Phi^{\mu\nu} - B^2 \nonumber\\
&-& B \left(\partial_\mu \phi^\mu + m \,\varphi \right) + B_\mu B^\mu - B^\mu \left(\partial^\nu B_{\nu\mu} - \partial_\mu \varphi + m \phi_\mu \right)
- m^2\, \bar \beta \beta \nonumber\\ 
&+& \left(\partial_\mu \bar C_\nu - \partial_\nu \bar C_\mu \right) \left(\partial^\mu C^\nu \right)
- \left(\partial_\mu \bar C - m \bar C_\mu \right) \left(\partial^\mu C - m C^\mu \right) + \partial_\mu \bar \beta \,\partial^\mu \beta  \nonumber\\
&+& \left(\partial_\mu \bar C^\mu + \frac{1}{2}\,\rho +  m \, \bar C\right) \lambda + \left(\partial_\mu C^\mu - \frac{1}{2}\,\lambda +  m \, C \right) \rho, 
\end{eqnarray}
\begin{eqnarray}
{\cal L}_{\bar B} &=& \frac{1}{12} \, H_{\mu\nu\eta}H^{\mu\nu\eta}
-\frac{1}{4}\,m^2\, B_{\mu\nu}B^{\mu\nu} -\frac{1}{4}\, \Phi_{\mu\nu} \Phi^{\mu\nu} + \frac{1}{2}\, m\, B_{\mu\nu} \Phi^{\mu\nu} - {\bar B}^2 \nonumber\\
&+& \bar B \left(\partial_\mu \phi^\mu - m \,\varphi\right) + \bar B_\mu \bar B^\mu 
+ \bar B^\mu \left(\partial^\nu B_{\nu\mu} + \partial_\mu \varphi + m \phi_\mu \right) - m^2\, \bar \beta \beta \nonumber\\ 
&+& \left(\partial_\mu \bar C_\nu - \partial_\nu \bar C_\mu \right)(\partial^\mu C^\nu)
-  \left(\partial_\mu \bar C - m \bar C_\mu \right) \left(\partial^\mu C - m C^\mu \right)  + \partial_\mu \bar \beta\, \partial^\mu \beta \nonumber\\
&+& \left(\partial_\mu \bar C^\mu + \frac{1}{2}\,\rho +  m \, \bar C\right) \lambda + \left(\partial_\mu C^\mu - \frac{1}{2}\,\lambda + m \, C\right) \rho, 
\end{eqnarray}
where the vector fields ${\bar B}_\mu,\; B_\mu$ and scalar fields $\bar B,\;B$ are the 
Nakanishi--Lautrup type  auxiliary fields, the vector fields
$(\bar C_\mu)C_\mu$ and scalar fields $(\bar C)C$  (with ${\bar C}_\mu \bar C^\mu = C_\mu C^\mu =0$, $C_\mu\bar C_\nu 
+ {\bar C}_\nu C_\mu =0$, $C_\mu C_\nu + {C}_\nu C_\mu =0$, ${\bar C}^2 = C^2 =0$, $C \bar C + {\bar C} C =0$, etc.) are the
fermionic (anti-)ghost fields, $\bar\beta$, $\beta$ are the bosonic ghost-for-ghost fields, ($\rho)\lambda$ are the fermionic auxiliary 
(anti-)ghost fields. The  fermionic (anti-)ghost fields $(\bar C_\mu)C_\mu$,  $(\bar C)C$ and ($\rho)\lambda$ carry  ghost number equal to ($-1)+1$ 
whereas bosonic (anti-)ghost fields  ($\bar\beta)\beta$ have ghost number equal to ($-2)+2$. The remaining fields carry zero ghost number. 
The commuting (anti-)ghost fields $(\bar \beta)\beta$  and  scalar field $\varphi$ 
are required for the stage-one reducibility in the theory (see, e.g.  \cite{mb} for details).

The above Lagrangian densities respect the following off-shell nilpotent (i.e. $s^2_{(a)b} = 0$) and absolutely anticommuting
(i.e. $s_b \, s_{ab} + s_{ab}\, s_b = 0$) (anti-)BRST symmetry transformations $(s_{(a)b})$:  
\begin{eqnarray}
&&  s_b B_{\mu\nu} = - (\partial_\mu C_\nu - \partial_\nu C_\mu), \qquad 
s_b C_\mu  = - \partial_\mu \beta,  \qquad s_b \phi_\mu = \partial_\mu C - m\, C_\mu, \nonumber\\
&& s_b \bar C_\mu = B_\mu, \quad s_b \bar \beta = - \rho, \quad s_b C = - m\,\beta, \quad s_b \bar C =  B, \quad s_b \bar B = - m \lambda,  \nonumber\\
&& s_b \varphi = \lambda, \qquad s_b \bar B_\mu =  - \partial_\mu \lambda, \qquad s_b [B, \rho, \lambda, \beta, B_\mu, H_{\mu\nu\kappa}] = 0,
\end{eqnarray}
\begin{eqnarray}
&&  s_{ab} B_{\mu\nu} = - (\partial_\mu \bar C_\nu - \partial_\nu \bar C_\mu), \qquad 
s_{ab} \bar C_\mu  = - \partial_\mu \bar \beta, \qquad s_{ab} \phi_\mu = \partial_\mu \bar C - m\, \bar C_\mu, \nonumber\\
&&  s_{ab}  C_\mu =  \bar B_\mu, \quad s_{ab} \beta = - \lambda, \quad s_{ab} \bar C = - m\, \bar \beta, \quad  s_{ab}  C = \bar B, 
\quad s_{ab} B = - m \rho, \nonumber\\
&& s_{ab} \varphi = \rho, \qquad s_{ab} B_\mu =  - \partial_\mu \rho, \qquad s_{ab} [\bar B, \rho, \lambda, \bar \beta, \bar B_\mu,  H_{\mu\nu\kappa}] = 0,
\end{eqnarray}
It is straightforward  to check that the Lagrangian densities ${\cal L}_B$ and ${\cal L}_{\bar B}$  
under the off-shell nilpotent BRST and anti-BRST  symmetry transformations transform to the 
total spacetime derivatives, respectively, as 
\begin{eqnarray}
s_b {\cal L}_B &=& - \partial_\mu \, \Big[B (\partial^\mu C - m C^\mu) - B_\nu (\partial^\mu C^\nu-\partial^\nu C^\mu) 
+ \rho (\partial^\mu \beta) - \lambda\, B^\mu \Big], \nonumber\\ 
s_{ab} {\cal L}_{\bar B} &=& \partial_\mu  \Big[\bar B (\partial^\mu \bar C + m \bar C^\mu) 
- \bar B_\nu (\partial^\mu \bar C^\nu - \partial^\nu \bar C^\mu) 
- \lambda (\partial^\mu \bar \beta) + \rho \bar B^\mu \Big].
\end{eqnarray}
As a consequence,  the action integrals 
remain invariant (i.e. $ s_b \int dx^4 {\cal L}_B =0, \; s_{ab}\int dx^4 {\cal L}_{\bar B} =0$) 
under the nilpotent (anti-)BRST transformations (10) and (11).

At this juncture,  the following remarks are in order:
\begin{enumerate}[label = $(\roman{*})$]
\item The above Lagrangian densities are coupled because the Nakanishi--Lautrup type auxiliary fields $B, \bar B$ and 
$B_\mu,\, \bar B_\mu $ are related to each other through the celebrated Curci--Ferrari (CF)
type of conditions:
\begin{eqnarray}
&& B + \bar B + m \varphi = 0, \qquad B_\mu + \bar B_\mu + \partial_\mu \varphi = 0.
\end{eqnarray} 

\item It is to be noted that ${\cal L}_B$ and ${\cal L}_{\bar B}$ transform under the continuous anti-BRST and BRST 
transformations, respectively, as follows: 
\begin{eqnarray}
s_{ab} {\cal L}_B &=& \partial_\mu \, \Big[ B_\nu (\partial^\mu \bar C^\nu - \partial^\nu \bar C^\mu)
- B (\partial^\mu \bar C - m \bar C^\mu) - \lambda (\partial^\mu \bar \beta) \nonumber\\
&+& \rho \left(\partial_\nu B^{\nu \mu} + \bar B^\mu +  m\, \phi^\mu \right)\Big]
+  m \rho \Big[B + \bar B + m \varphi \big] - (\partial_\mu \rho)\Big[B^\mu + \bar B^\mu + \partial^\mu \varphi \Big]  \nonumber\\
&-& m \Big[B^\mu + \bar B^\mu + \partial^\mu \varphi \Big](\partial^\mu \bar C - m \bar C^\mu)
+ \partial_\mu \Big[B + \bar B + m \, \varphi\Big] (\partial^\mu \bar C - m\,\bar C^\mu) \nonumber\\
&+& \partial_\mu \Big[B_\nu + \bar B_\nu + \partial_\nu \varphi \Big] (\partial^\mu \bar C^\nu - \partial^\nu \bar C^\mu), \nonumber\\
&&\nonumber\\
s_b {\cal L}_{\bar B} &=& \partial_\mu \, \Big[\bar B (\partial^\mu C - m C^\mu) 
- \bar B_\nu (\partial^\mu  C^\nu - \partial^\nu  C^\mu) - \rho (\partial^\mu  \beta) \nonumber\\
&-& \lambda \left(\partial_\nu B^{\nu \mu} - B^\mu + m\, \phi^\mu \right)  \Big]
+ m  \lambda \Big[B + \bar B + m \varphi \big]
- (\partial_\mu \lambda)\Big[B^\mu + \bar B^\mu + \partial^\mu \varphi \Big] \nonumber\\ 
&+& m \Big[B^\mu + \bar B^\mu + \partial^\mu \varphi \Big](\partial^\mu C - m C^\mu)
- \partial_\mu \Big[B + \bar B + m \, \varphi\Big] (\partial^\mu C - m\,C^\mu) \nonumber\\
&+& \partial_\mu \Big[B_\nu + \bar B_\nu + \partial_\nu \varphi \Big] (\partial^\mu  C^\nu - \partial^\nu C^\mu),
\end{eqnarray}
As a consequence, the coupled Lagrangian densities respect both BRST and anti-BRST symmetries on the 4D constraints hypersurface defined by the 
CF conditions (13). This reflects the fact that the coupled Lagrangian densities are equivalent on the constrained hypersurface 
defined by CF type of restrictions.  

\item The CF conditions are (anti-)BRST invariant as one can check that 
\begin{eqnarray}
&& s_{(a)b}\big[B + \bar B + m\, \varphi \big] = 0, \nonumber\\
&& s_{(a)b}\big[B_\mu + \bar B_\mu + \partial_\mu \varphi \big] = 0.
\end{eqnarray}
Thus, these conditions are physical conditions.

\item Further, the absolute anticommutativity property of the (anti-)BRST symmetry transformations is satisfy due the existence of the 
CF conditions. For the sake of brevity, we note that
\begin{eqnarray}
\{s_b, \, s_{ab}\}B_{\mu\nu} &=& - \partial_\mu(B_\nu + \bar B_\nu) + \partial_\nu(B_\mu + \bar B_\mu),\nonumber\\
\{s_b, \, s_{ab}\}\Phi_\mu &=& + \partial_\mu(B + \bar B) - m\, (B_\mu + \bar B_\mu).
\end{eqnarray}     
Now, it is clear from the above that $\{s_b, \, s_{ab}\}B_{\mu\nu} = 0$ and  $\{s_b, \, s_{ab}\}\phi_\mu =0$ if and only if the CF conditions are satisfied.
For the remaining fields, the anticommutativity property is trivially satisfied. 
\end{enumerate}  
To sum up the above results, we again emphasize on the fact that the CF conditions play a decisive role in providing 
the absolute anticommutativity of the (anti-)BRST transformations. 
These are also responsible for the coupled (but equivalent) Lagrangian densities. Furthermore, the CF type conditions are the 
physical restrictions (on the theory) in the sense that they are (anti-)BRST invariant conditions. We shall see later on that these CF conditions emerge 
very naturally within the framework of superfield approach to BRST formalism (cf. Sect. 6, below).


\section{Conserved (anti-)BRST Charges}


According to Noether's theorem, the invariance of the actions (corresponding to the coupled Lagrangian densities)
under the continuous (anti-)BRST symmetries yield the conserved (anti-)BRST currents $J^\mu_{(a)b}$:      
\begin{eqnarray}
J^\mu_b&=& - \frac{1}{2}\big(\partial_\nu C_\eta - \partial_\eta C_\nu\big) H^{\mu\nu\eta} 
+ B_\nu \big(\partial^\mu C^\nu - \partial^\nu C^\mu\big) - B \big(\partial^\mu C - m\, C^\mu\big) \nonumber\\
&-& \big(\partial_\nu C - m \, C_\nu\big) \big(\Phi^{\mu\nu} - m \, B^{\mu\nu}\big) 
+ \big(\partial^\mu \bar C^\nu - \partial^\nu \bar C^\mu\big) (\partial_\nu \beta) \nonumber\\
&-& m\, \beta \big(\partial^\mu \bar C - m\, \bar C^\mu\big) - \rho \,(\partial^\mu \beta) + \lambda\, B^\mu, \nonumber\\
&&\nonumber\\
J^\mu_{ab} & = & - \frac{1}{2}\big(\partial_\nu \bar C_\eta - \partial_\eta \bar C_\nu\big) H^{\mu\nu\eta} 
- \bar B_\nu \big(\partial^\mu \bar C^\nu - \partial^\nu \bar C^\mu\big) + \bar B \big(\partial^\mu \bar C - m\, \bar C^\mu\big) \nonumber\\
&-& \big(\partial_\nu \bar C - m \, \bar C_\nu\big) \big(\Phi^{\mu\nu} - m \, B^{\mu\nu}\big) 
- \big(\partial^\mu C^\nu - \partial^\nu C^\mu\big) (\partial_\nu \bar \beta)  \nonumber\\
&+&  m\, \bar \beta \big(\partial^\mu C - m\, C^\mu\big) - \lambda \,(\partial^\mu \bar \beta) + \rho\, \bar B^\mu. 
\end{eqnarray}
The conservation $(\partial_\mu J^\mu_b = 0)$ of BRST current $J^\mu_b$ can be proven by using the 
following Euler--Lagrange equations of motion:   
\begin{eqnarray}
&& \partial_\mu H^{\mu\nu\eta} - \big(\partial^\nu B^\eta - \partial^\eta B^\nu\big) - m \big(\Phi^{\nu\eta} - m  B^{\nu\eta}\big) = 0, \nonumber\\
&& \partial_\mu \Phi^{\mu\nu} + \partial^\nu B - m \big(\partial_\mu B^{\mu\nu} + B^\nu \big) = 0, 
\qquad B_\mu = \frac{1}{2}\, \big(\partial^\nu B_{\nu\mu} - \partial_\mu \varphi + m \phi_\mu\big),\nonumber\\
&& B = - \, \frac{1}{2}\, \big(\partial_\mu \phi^\mu + m \varphi\big), \qquad \partial_\mu B^\mu + m \, B = 0, \nonumber\\
&&\Box C_\mu - \partial_\mu (\partial_\nu C^\nu) + \partial_\mu \lambda - m \big(\partial_\mu C - m \, C_\mu\big) = 0, 
\qquad  \Box C - m (\partial_\nu C^\nu) + m \, \lambda = 0,\nonumber\\
&& \Box \bar C_\mu - \partial_\mu (\partial_\nu \bar C^\nu) - \partial_\mu \rho - m \big(\partial_\mu \bar C - m \, \bar C_\mu\big) = 0,
\qquad \Box \bar C - m (\partial_\nu \bar C^\nu) - m \, \rho = 0,  \nonumber\\
&& \big(\Box + m^2 \big)\beta = 0, \qquad  \lambda = \big(\partial_\mu C^\mu + m\, C\big), \nonumber\\
&& \big(\Box + m^2 \big) \bar \beta = 0, \qquad \rho = - \big(\partial_\mu \bar C^\mu + m\, \bar C\big). 
\end{eqnarray}
These equations of motion have been derived from ${\cal L}_B$. Similarly, for the conservation 
$(\partial_\mu J^\mu_{ab} = 0)$ of anti-BRST current $J^\mu_{ab}$, we have used the equations of motion derived from ${\cal L}_{\bar B}$. 
We point out that most of the equations of motion are the same for ${\cal L}_{B}$ and ${\cal L}_{\bar B}$. The Euler--Lagrange equations of motion that 
are different from (18) and derived from ${\cal L}_{\bar B}$ are listed as follows:
\begin{eqnarray}
&& \partial_\mu H^{\mu\nu\eta} + \big(\partial^\nu \bar B^\eta - \partial^\eta \bar B^\nu\big) 
- m \big(\Phi^{\nu\eta} - m \, B^{\nu\eta}\big) = 0, \nonumber\\
&& \partial_\mu \Phi^{\mu\nu} - \partial^\nu \bar B - m \big(\partial_\mu B^{\mu\nu} - \bar B^\nu \big) = 0, 
\qquad  \bar B_\mu = -\,\frac{1}{2}\, \big(\partial^\nu B_{\nu\mu} + \partial_\mu \varphi + m \phi_\mu\big),\nonumber\\
&& \bar B = \frac{1}{2}\, \big(\partial_\mu \phi^\mu - m \varphi\big), \qquad \partial_\mu \bar B^\mu + m \, \bar B = 0.
\end{eqnarray}
It is interesting to mention that the appropriate equations of motion  derived from ${\cal L}_B$ and ${\cal L}_{\bar B}$ [cf. (18) and (19)] 
produce the CF conditions (13).

The temporal components of the conserved currents  (i.e. $Q_{(a)b} = \int d^3x J^0_{(a)b}$) lead to the following 
charges $Q_{(a)b}$: 
\begin{eqnarray}
Q_b&=& \int d^3x \Big[- \frac{1}{2}\big(\partial_i C_j - \partial_j C_i\big) H^{0ij} 
+ B_i \big(\partial^0 C^i - \partial^i C^0\big) - B \big(\partial^0 C - m C^0\big)\nonumber\\
&-& \big(\partial_i C - m  C_i\big) \big(\Phi^{0i} - m  B^{0i}\big) - m \beta \big(\partial^0 \bar C - m \bar C^0\big) \nonumber\\
&+& \big(\partial^0 \bar C^i - \partial^i \bar C^0\big) (\partial_i \beta) - \rho \,(\partial^0 \beta) + \lambda\, B^0 \Big],\nonumber\\
&&\nonumber\\
Q_{ab} & = & \int d^3x \Big[- \frac{1}{2}\big(\partial_i \bar C_j - \partial_j \bar C_i\big) H^{0ij} 
-\bar B_i \big(\partial^0 \bar C^i - \partial^i \bar C^0\big) + \bar B \big(\partial^0 \bar C - m \bar C^0\big) \nonumber\\
&-& \big(\partial_i \bar C - m  \bar C_i\big) \big(\Phi^{0i} - m  B^{0i}\big) + m \bar \beta \big(\partial^0 C - m\, C^0\big)\nonumber\\
&-& \big(\partial^0 C^i - \partial^i C^0\big) (\partial_i \bar \beta) - \lambda \,(\partial^0 \bar \beta) + \rho\, \bar B^0 \Big]
\end{eqnarray}
It turns out that these conserved charges are the generators of the corresponding symmetry transformations. For instance, one can check that 
\begin{eqnarray}
&& s_{(a)b} \Psi = - i \big[\Psi, Q_{(a)b}\big]_{\pm},  \qquad \Psi = B_{\mu\nu},\phi_\mu, C_\mu, \bar C_\mu, \beta, \bar \beta, C, \bar C, \varphi, 
\end{eqnarray}
where $(\pm)$ signs, as the subscript, on the square bracket correspond to the (anti)commutator  depending on the generic field 
$\Psi$ being (fermionic)bosonic in nature. We, further, point out that the conserved (anti-)BRST charges do not produce the proper (anti-)BRST
symmetry  transformations for the Nakanishi--Lautrup type 
auxiliary fields $B$, $\bar B$, $B_\mu$, $\bar B_\mu$ and the auxiliary (anti-) ghost fields $(\rho)\lambda$. 
The transformations of these auxiliary fields
can be derived from the requirements of the nilpotency and absolute anticommutativity properties of the (anti-)BRST symmetry transformations.

The (anti-)BRST charges are nilpotent and anticommuting in nature. These properties can be shown in a straightforward manner by exploiting 
the definition of a generator. For the nilpotency of the (anti-)BRST charges, the following relations are true: 
\begin{eqnarray}
s_b Q_b &=& - i \big\{Q_b,\, Q_b \big\} = 0  \Rightarrow Q^2_b = 0, \nonumber\\
s_{ab} Q_{ab} &=& - i \big\{Q_{ab},\, Q_{ab} \big\} = 0  \Rightarrow Q^2_{ab} = 0.
\end{eqnarray} 
In a similar fashion, one can also show the anticommutativity of the (anti-)BRST charges as
\begin{eqnarray}
s_b Q_{ab} = - i \big\{Q_{ab}, Q_b \big\} = 0 \Rightarrow Q_b Q_{ab} + Q_{ab} Q_b = 0,\nonumber\\
s_{ab} Q_b = - i \big\{Q_b, Q_{ab} \big\} = 0 \Rightarrow Q_b Q_{ab} + Q_{ab} Q_b = 0.
\end{eqnarray}
The above computations are more algebraically involved. For the shake of completeness,  in our Appendix  A, we shall provide a complete proof 
of the first relation that appear in (23) in a simpler way.

Before we wrap up this section, we dwell a bit on the constraint structure of the gauge-invariant Lagrangian density (5) within the 
framework of BRST formalism.  We define a physical state $(|phys \rangle)$ in the quantum Hilbert space of states which respects the 
(anti-) BRST symmetries. The physicality criteria $Q_{(a)b} \, | phys\rangle  = 0$ state that the physical state $|phys \rangle$ must be 
annihilated by the conserved and nilpotent (anti-)BRST charges $Q_{(a)b}$. In other words, we can say that Faddeev-Popov ghosts are 
decoupled from the physical states of the theory. Thus, the physicality criterion $Q_b \, | phys\rangle  = 0$
produces the following constraint conditions:   
\begin{eqnarray}
- B\,|phys\rangle &=& 0, \nonumber\\
\partial_i \big(\Phi^{0i} - m B^{0i} \big) |phys \rangle &=& 0, \nonumber\\
 B^i \,|phys\rangle &=& 0, \nonumber\\
-\big(\partial_j H^{0ij} + m (\Phi^{0i} - m B^{0i}) \big)\, |phys \rangle &=&0,
\end{eqnarray}
which, finally, imply the familiar constraint conditions on the physical state: $\Pi^0 |phys\rangle = 0$,
 $\partial_i \Pi^i |phys\rangle = 0$,  $\Pi^{0i} |phys \rangle = 0$,
 $-\big(2 \partial_j \Pi^{ij} + m \, \Pi^i\big) |phys\rangle = 0$,
where  $\Pi^0$, $\Pi^i$, $\Pi^{0i}$, $\Pi^{ij}$ are the canonical conjugate momenta with respect to the dynamical fields 
$\phi_0$, $\phi_i$, $B_{0i}$, $B_{ij}$, respectively. These momenta have been derived from the Lagrangian density (8).  The very 
similar constraint conditions also emerge when we exploit the physicality criterion $Q_{ab}\,|phys \rangle = 0$. These constraint 
conditions are consistent with gauge-invariant Lagrangian (5). As a consequence, the BRST quantization is consistent with the 
requirements of the Dirac quantization scheme for the constrained systems.


\section{Ghost-scale symmetry and BRST algebra}

The coupled Lagrangian densities, in addition to the (anti-)BRST symmetries, also respect the following continuous ghost-scale symmetry 
transformations: 
\begin{eqnarray}
&& C_\mu \to e^{+ \vartheta} \, C_\mu, \qquad \bar C_\mu \to e^{- \vartheta} \, \bar C_\mu, 
\qquad C \to e^{+ \vartheta} \, C, \qquad \bar C \to e^{- \vartheta} \, \bar C,    \nonumber\\
&& \beta \to e^{+2 \vartheta} \, \beta, \qquad \bar \beta \to e^{-2 \vartheta} \, \bar \beta, \qquad  \lambda \to e^{+ \vartheta} \, \lambda, 
\qquad  \rho \to e^{-  \vartheta} \, \rho,\nonumber\\
&& (B_{\mu\nu}, \phi_\mu, B_\mu, \bar B_\mu, B, \bar B, \varphi) \to e^0\, (B_{\mu\nu}, \phi_\mu, B_\mu, \bar B_\mu, B, \bar B, \varphi),
\end{eqnarray}
where $\vartheta$ is a (spacetime independent) global scale parameter. The numerical factors in the exponentials 
(i.e. $0,\pm 1$, $\pm 2$) define the ghost number of the various fields present in the theory. The infinitesimal 
version of the above ghost-scale symmetry (with $\vartheta = 1$) leads to the following symmetry transformations ($s_g$):
\begin{eqnarray}
&& s_g C_\mu = + \, C_\mu, \qquad  s_g \bar C_\mu = - \, \bar C_\mu, \qquad  s_g C = + \, C, \qquad  s_g \bar C = - \, \bar C, \nonumber\\
&& s_g \beta = + 2 \, \beta, \qquad  s_g \bar \beta = - 2 \, \bar \beta, \qquad  s_g \rho = - \, \rho,  \qquad s_g \lambda = + \, \lambda, \nonumber\\
&& s_g \big[B_{\mu\nu}, \phi_\mu, B_\mu, \bar B_\mu,B, \bar B, \varphi \big] = 0,
\end{eqnarray}
under which the (coupled) Lagrangian densities remain invariant (i.e. $s_g {\cal L}_{B} = s_g {\cal L}_{\bar B} = 0$).

According to Noether theorem, the continuous ghost-scale symmetry yields the conserved current $J^\mu_g$ and corresponding charge $Q_g$, namely;
\begin{eqnarray}
J^\mu_g &=& \big(\partial^\mu \bar C^\nu - \partial^\nu \bar C^\mu \big)\,C_\nu 
+ \big(\partial^\mu C^\nu - \partial^\nu C^\mu \big)\, \bar C_\nu  
- \big(\partial^\mu \bar C - m\, \bar C^\mu\big)\, C - \big(\partial^\mu C - m\, C^\mu\big)\, \bar C  \nonumber\\
&+& 2 \beta (\partial^\mu \bar \beta) - 2 \bar \beta (\partial^\mu \beta) - \rho\, C^\mu + \lambda\, \bar C^\mu, \nonumber\\
&& \nonumber\\
Q_g &=& \big(\partial^0 \bar C^i - \partial^i \bar C^0 \big)\,C_i 
+ \big(\partial^0 C^i - \partial^i C^0 \big)\, \bar C_i  
- \big(\partial^0 \bar C - m\, \bar C^0\big)\, C - \big(\partial^0 C - m\, C^0 \big)\, \bar C  \nonumber\\
&+& 2 \beta \big(\partial^0 \bar \beta \big) - 2 \bar \beta \big(\partial^0 \beta \big) - \rho\, C^0 + \lambda\, \bar C^0. 
\end{eqnarray}
It is evident that the above charge is the generator of the corresponding ghost-scale symmetry transformations as one can check that 
\begin{eqnarray}
&& s_g \Psi = \pm \,i \big[\Psi, \, Q_g\big], 
\end{eqnarray}
where $\Psi$ is the generic field of the theory. The $(\pm)$ signs in front of the commutator are used for the generic field $\Psi$ being 
(fermionic) bosonic in nature.

At this moment, the following remarks are in order:
\begin{enumerate}[label = $(\roman{*})$]
\item The conserved ghost charge $Q_g$ does not produce the proper transformations for the auxiliary fields
$\rho$ and $\lambda$. These transformations can be obtained from other considerations (see (29) below).  

\item  The continuous symmetry transformations (in their operator form) obey the following algebra:
\begin{eqnarray}
&& s^2_b = 0, \qquad s^2_{ab} = 0, \qquad \big\{s_b, \, s_{ab}\big\} = 0, \qquad \big[s_g, s_g \big] = 0,\nonumber\\
&& \big[s_g, \, s_b \big] = + \, s_b, \qquad \big[s_g, \, s_{ab} \big] = - \, s_{ab}.
\end{eqnarray}

\item By exploiting the last two relations of the above equation, we can obtain the proper transformations for $\rho$ and $\lambda$.
For instance, one can check that
\begin{eqnarray}
\big[s_g, \, s_b \big] \beta = + \, s_b \bar \beta \Rightarrow s_g \rho = - \, \rho.
\end{eqnarray}
Similarly, the transformation
for the auxiliary field $\lambda$ can be derived, too.

\item The operator form of the conserved (anti-)BRST charges together with the ghost charge obeys the following graded algebra: 
\begin{eqnarray}
&& Q^2_b = 0, \qquad Q^2_{ab} = 0, \qquad \big\{Q_b, \, Q_{ab}\big\} = 0, \qquad \big[Q_g, Q_g] = 0\nonumber\\
&& \big[Q_g, \, Q_b \big] = - i \, Q_b, \qquad \big[Q_g, \, Q_{ab} \big] = + i \, Q_{ab},
\end{eqnarray} 
which is also known as standard BRST algebra.

As a consequences of the above algebra, 
we define an eigenstate $|\zeta\rangle_n$ (in the quantum Hilbert space of states) with respect to the operator $i Q_g$ 
such that $i Q_g |\zeta\rangle_n = n |\zeta\rangle_n$. Here $n$ defines the ghost number 
as the eigenvalue of the operator $i Q_g$. 
Using the above algebra among the conserved charges, it is straightforward to check that the following relationships are true: 
\begin{eqnarray}
i Q_g\, Q_b |\zeta\rangle_n &=& (n + 1)\, Q_b |\zeta\rangle_n, \nonumber\\
i Q_g\, Q_{ab} |\zeta\rangle_n &=& (n - 1)\, Q_{ab} |\zeta\rangle_n,
\end{eqnarray}   
which imply that the eigenstates $Q_b |\zeta\rangle_n$ and $Q_{ab} |\zeta\rangle_n$ have the eigenvalues $(n+1)$ and $(n -1)$, respectively.
In other words, The conserved (anti-)BRST  charges $Q_{(a)b}$ (decrease)increase the ghost number of the eigenstate $i Q_g |\zeta\rangle_n$ by one unit. 
Also, we can say that the (anti-)BRST charges $Q_{(a)b}$ carry ghost number equal to $(-1)+1$ while ghost charge $Q_g$ does not carry any
ghost number. These observations also reflect from the expressions of the conserved charges if we look carefully for
the ghost number of the various fields that appear in the charges.

\end{enumerate}


\section{Augmented superfield approach to BRST formalism}


In this section, we shall derive the {\it proper} off-shell nilpotent and absolutely anticommuting (anti-)BRST symmetry
transformations with the help of an extended version of Bonora--Tonin superfield formalism \cite{bt1,bt2} where the horizontality condition
and gauge-invariant restriction are used in a physically meaningful manner.  
In this formalism, we generalize our ordinary 4D spacetime to 
$(4,2)D$ superspace parameterized by an additional pair of the Grassmannian variables ($\theta, \bar \theta$) as
\begin{eqnarray}
x^\mu \to Z^M = (x^\mu, \theta, \bar \theta), \quad \partial_\mu \to \partial_M = (\partial_\mu , \partial_\theta, \partial_{\bar \theta}),
\end{eqnarray}
where $x^\mu$ ($\mu = 0, 1, 2, 3$) are the (bosonic) spacetime coordinates. The super coordinates $Z^M = (x^\mu, \theta, \bar \theta)$ 
parametrize the $(4, 2)$D superspace (with $\theta^2 = 0, \, \bar \theta^2 = 0, \, \theta \, \bar \theta + \bar \theta\, \theta = 0$) and 
$\partial_\theta = \partial/\partial \theta,$ $\partial_{\bar \theta} = \partial/\partial \bar \theta$ are the Grassmannian 
translational generators along the Grassmannian directions $\theta$, $\bar \theta$. We generalize the exterior derivative 
$d$ and 2-form  $B^{(2)}$  to the super exterior derivative $\tilde d$ and super 2-form $\tilde {\cal B}^{(2)}$ on the $(4, 2)$D 
supermanifold  as follows:      
\begin{eqnarray}
d \to \tilde d &=& d Z^M \partial_M \nonumber\\
&\equiv& dx^\mu \,\partial_\mu + d \theta\,\partial_\theta + d \bar\theta \partial_{\bar\theta}, 
\qquad \tilde d^2 = 0, 
\end{eqnarray}
\begin{eqnarray}
B^{(2)}\to \tilde {\cal B}^{(2)} &=& {\displaystyle \frac{1}{2!}} \big(dZ^M \wedge dZ^N \big) \, \tilde {\cal B}_{MN}\nonumber\\ 
&\equiv& {\displaystyle \frac{1}{2!}} \big(dx^\mu \wedge dx^\nu \big) \, \tilde {\cal B}_{\mu\nu} (x, \theta, \bar\theta) \nonumber\\
&+& \big(dx^\mu \wedge d \theta \big)\, {\tilde {\bar {\cal F}}}_\mu (x, \theta, \bar\theta) 
+ \big(dx^\mu \wedge d \bar\theta \big)\,{\tilde {\cal F}}_\mu (x, \theta, \bar\theta) 
+ \big(d\theta \wedge d \theta \big)\, {\tilde {\bar \beta}} (x, \theta, \bar\theta) \nonumber\\
&+& \big(d \bar\theta \wedge d \bar\theta \big) \, {\tilde \beta}(x, \theta, \bar\theta) 
+ \big(d \theta \wedge d\bar\theta \big)\, {\tilde \Phi} (x, \theta, \bar\theta).
\end{eqnarray}
The super multiplets as the components of the super 2-from can be expanded along the Grassmannian directions 
($\theta$  and $\bar \theta$) as follows:    
\begin{eqnarray}
{\tilde {\cal B}_{\mu\nu}} (x, \theta, \bar\theta) &=& B_{\mu\nu}(x) + \theta\, \bar R_{\mu\nu} (x) + \bar\theta\, R_{\mu\nu} (x) 
+ i \,\theta \, \bar\theta\, S_{\mu\nu} (x), \nonumber\\ 
{\tilde {\cal F}}_\mu (x, \theta, \bar\theta) &=& C_\mu (x) + \theta \,\bar B^{(1)}_\mu (x) + \bar\theta\, B^{(1)}_\mu (x)
+ i \,\theta \, \bar\theta\,f_\mu (x), \nonumber\\ 
{\tilde {\bar {\cal F}}}_\mu (x, \theta, \bar\theta) &=& \bar C_\mu (x) + \theta\, \bar B^{(2)}_\mu (x) + \bar\theta\, B^{(2)}_\mu (x) 
+ i \, \theta\, \bar\theta \bar f_\mu (x), \nonumber\\
{\tilde \beta} (x, \theta, \bar\theta ) &=& \beta (x) + \theta \,\bar f_1 (x) + \bar\theta\, f_1 (x) + i\, \theta\, \bar\theta\, b_1 (x), \nonumber\\ 
{\tilde {\bar \beta}} (x, \theta, \bar\theta) &=& \bar\beta (x) + \theta \,\bar f_2 (x) + \bar \theta\, f_2 (x)+ i \,\theta\,\bar\theta\, b_2 (x), \nonumber\\ 
{\tilde \Phi} (x,\theta, \bar\theta) &=& \varphi (x) + \theta \,\bar f_3 (x) + \bar\theta\, f_3 (x) + i \,\theta \,\bar\theta\, b_3 (x),
\end{eqnarray}
where the secondary fields $R_{\mu\nu}$, $\bar R_{\mu\nu}$, $f_\mu$, $\bar f_\mu$, $f_1$, $\bar f_1$, $f_2$,  $\bar f_2$, $f_3$, $\bar f_3$ 
are fermionic in nature  and $S_{\mu\nu}$, $B^{(1)}_\mu$,  $\bar B^{(1)}_\mu$, $B^{(2)}_\mu$, $\bar B^{(2)}_\mu$, $b_1$, $b_2$, $b_3$ 
are bosonic secondary fields. We shall determine the precise value of these secondary fields with the help of superfield formalism.

It is to be noted that the following horizontality condition (HC), 
\begin{eqnarray}
&& d B^{(2)} = \tilde d \tilde {\cal B}^{(2)} \Longleftrightarrow  H^{(3)} = \tilde {\cal H}^{(3)},
\end{eqnarray}
determines the value of all secondary fields in terms of the basic and auxiliary fields of the theory. 
This HC implies that the l.h.s. is independent of the Grassmannian variables $\theta$ and $\bar \theta$ when we 
generalize it on the $(4,2)$D supermanifold. The r.h.s. of (37), in its full blaze of glory, can be written as
\begin{eqnarray}
\tilde {\cal H}^{(3)} &=& \tilde d \tilde {\cal B}^{(2)}\nonumber\\
 &=& {\displaystyle \frac{1}{3!}}
(dx^\mu \wedge d x^\nu \wedge dx^\kappa) \big(\partial_\mu \tilde {\cal B}_{\nu\kappa} + \partial_\nu \tilde {\cal B}_{\kappa\mu}
+ \partial_\kappa \tilde {\cal B}_{\mu\nu} \big) \nonumber\\
&+& {\displaystyle \frac{1}{2!}} (dx^\mu \wedge dx^\nu \wedge d \theta) \bigl [\partial_{\theta} \tilde {\cal B}_{\mu\nu} 
+ \partial_\mu \tilde {\bar {\cal F}}_\nu -\partial_\nu \tilde {\bar {\cal F}}_\mu \bigr ] \nonumber\\
&+& {\displaystyle \frac{1}{2!}} (dx^\mu \wedge dx^\nu \wedge d \bar \theta) \bigl [\partial_{\bar\theta} \tilde {\cal B}_{\mu\nu} 
+ \partial_\mu \tilde {\cal F}_\nu - \partial_\nu \tilde  {\cal F}_\mu \bigr ] \nonumber\\
&+& (d \theta \wedge d \theta \wedge d \theta) \big(\partial_{\theta} \tilde {\bar \beta} \big)
+ (d \bar\theta \wedge d \bar \theta \wedge d \bar\theta) \big(\partial_{\bar\theta} \tilde\beta \big) \nonumber\\ 
&+& (dx^\mu \wedge d\theta \wedge d\bar\theta) \bigl [\partial_\mu \tilde \Phi + \partial_\theta \tilde {\cal F}_\mu 
+\partial_{\bar\theta} \tilde {\bar {\cal F}}_\mu \bigr ]\nonumber\\
&+& (dx^\mu \wedge d \theta \wedge d \theta) \bigl [\partial_{\theta} {\bar {\cal F}}_\mu + \partial_\mu \tilde {\bar \beta} \bigr ] 
+ (dx^\mu \wedge d \bar\theta \wedge d \bar \theta) \bigl [\partial_{\bar\theta} \tilde {\cal F}_\mu + \partial_\mu \tilde  \beta \bigr ]\nonumber\\
&+& (d\theta \wedge d\bar\theta \wedge d\bar\theta) \bigl [\partial_{\bar\theta} \tilde \Phi + \partial_\theta \tilde \beta \bigr ] 
+ (d\bar \theta \wedge d\theta \wedge d\theta) \bigl [\partial_{\theta} \tilde \Phi +\partial_{\bar\theta} \tilde {\bar\beta} \bigr ].
\end{eqnarray}
The above HC implies the following interesting relationships amongst the superfields:     
\begin{eqnarray}
&& \partial_\theta {\tilde {\cal B}_{\mu\nu}} + \partial_\mu {\tilde {\bar {\cal F}}}_\nu -  \partial_\nu {\tilde {\bar {\cal F}}}_\mu  = 0, 
\qquad  \partial_{\bar\theta} {\tilde {\cal B}_{\mu\nu}} + \partial_\mu {\tilde { {\cal F}}}_\nu -  \partial_\nu {\tilde  {\cal  F}}_\mu  =0,
 \nonumber\\
&& \partial_\mu {\tilde { \Phi}} + \partial_\theta {\tilde {{\cal F}}}_\mu +  \partial_{\bar\theta} {\tilde {\bar {\cal F}}}_\mu  =0,  
\qquad \partial_\mu {\tilde { \bar\beta}} 
+ \partial_\theta {\tilde {\bar {\cal F}}}_\mu  =0, \qquad
\partial_\mu {\tilde { \beta}} 
+ \partial_{\bar\theta} {\tilde { {\cal F}}}_\mu  =0, \nonumber\\
&& \partial_\theta{\tilde { \Phi}} 
+ \partial_{\bar\theta} {\tilde {{\cal \bar\beta}}}  =0, \quad\qquad
\partial_{\bar\theta}{\tilde { \Phi}} 
+ \partial_{\theta} {\tilde {{\cal \beta}}}  =0, \qquad \partial_{\theta} {\tilde {{\cal\bar \beta}}} = 0, \qquad \partial_{\bar\theta} {\tilde {{\cal \beta}}} =0.
 \end{eqnarray}
Exploiting the above expressions for the superfields given in (36), we obtain the value of secondary fields, 
\begin{eqnarray}
&& R_{\mu\nu} = - (\partial_\mu C_\nu - \partial_\mu C_\mu), \qquad \bar R_{\mu\nu} = - (\partial_\mu \bar C_\nu - \partial_\mu \bar C_\mu), 
\qquad \bar B^{(2)} = - \partial_\mu \bar \beta,   \nonumber\\
&& S_{\mu\nu} = i \big(\partial_\mu B^{(2)}_\nu - \partial_\nu B^{(2)}_\mu \big) 
\equiv - i \big(\partial_\mu \bar B^{(1)}_\nu - \partial_\nu \bar B^{(1)}_\mu \big), \qquad \quad B^{(1)} = - \partial_\mu \beta,\nonumber\\ 
&& f_\mu = i\, \partial_\mu f_3 \equiv - i\,\partial_\mu \bar f_1, \qquad \bar f_\mu = -i\, \partial_\mu \bar f_3 
\equiv i\, \partial_\mu f_2,  \nonumber\\
&& B^{(2)}_\mu + \bar B^{(1)}_\mu + \partial_\mu \varphi = 0, \qquad  f_2 + \bar f_3 = 0, \qquad \bar f_2 + f_3 = 0,   \nonumber\\
&&  b_1 = 0, \qquad b_2 =0, \qquad b_3 = 0, \qquad f_1 = 0, \qquad \bar f_2 = 0.
\end{eqnarray}
Substituting these values in the expressions of the superfields (36), we have the following super-expansions:  
\begin{eqnarray}
{\tilde {\cal B}_{\mu\nu}}^{(h)} (x, \theta, \bar\theta) &=& B_{\mu\nu}(x) 
- \theta\, \big(\partial_\mu \bar C_\nu - \partial_\nu \bar C_\mu \big) (x) 
- \bar\theta\, \big(\partial_\mu C_\nu - \partial_\nu C_\mu \big)  (x)  \nonumber\\
&+& \theta \, \bar\theta\, \big(\partial_\mu B_\nu - \partial_\nu B_\mu\big) (x), \nonumber\\ 
{\tilde {\cal F}}^{(h)}_\mu (x, \theta, \bar\theta) &=& C_\mu (x) + \theta \,\bar B_\mu (x) - \bar\theta\, \big(\partial_\mu \beta \big) (x) 
- \theta \, \bar\theta\, \big(\partial_\mu \lambda \big) (x), \nonumber\\ 
{\tilde {\bar {\cal F}}}^{(h)}_\mu (x, \theta, \bar\theta) &=& \bar C_\mu (x) - \theta\, \big(\partial_\mu \bar \beta \big) (x) 
- \bar\theta\, B_\mu (x) + \theta\, \bar\theta \big(\partial_\mu \rho\big)(x), \nonumber\\
{\tilde \beta}^{(h)} (x, \theta, \bar\theta ) &=& \beta (x) - \theta \, \lambda (x), \nonumber\\ 
{\tilde {\bar \beta}}^{(h)} (x, \theta, \bar\theta) &=& \bar\beta (x)  - \bar \theta\, \rho (x), \nonumber\\ 
{\tilde \Phi}^{(h)} (x,\theta, \bar\theta) &=& \varphi (x) + \theta \,\rho(x) + \bar\theta\, \lambda (x). 
\end{eqnarray}
The superscript $(h)$ on the superfields denotes the expansion of the superfields obtained after the application of HC.
In the above super-expansions,  we have chosen  $\bar f_3 = \rho = - f_2, \; f_3 = \lambda = - \bar f_1, \;\bar B^{(1)}_\mu = \bar B_\mu,
B^{(2)}_\mu =   B_\mu$ where $B_\mu$ and $\bar B_\mu$ play the role of Nkanishi--Lautrup type auxiliary fields. 
These auxiliary fields are required for the linearization of the gauge-fixing terms as well as for the off-shell 
nilpotency of the (anti-)BRST symmetry transformations.

It is clear from the above super-expansions of the superfields
that the coefficients of $\bar \theta$ are the BRST transformations whereas the  coefficients of $\theta$
are the anti-BRST transformations. To be more precise,  the BRST transformation $(s_b)$ for any generic field $\Psi(x)$ is equivalent to the 
translation of the corresponding superfield $\tilde \Psi^{(h)}(x, \theta, \bar \theta)$ along the $\bar \theta$-direction while keeping $\theta$-direction
fixed. Similarly, the anti-BRST transformation $(s_{ab})$ can be obtained by taking the translation of the superfield along the $\theta$-direction while 
$\bar \theta$-direction remains intact. Mathematically, these statements can be corroborated in the following fashion:    
\begin{eqnarray}
&& s_b \Psi(x) = \frac{\partial}{\partial \bar \theta}\, \tilde \Psi^{(h)}(x, \theta, \bar \theta)\Big|_{\theta = 0}, \qquad 
s_{ab} \Psi(x) = \frac{\partial}{\partial \theta}\, \tilde \Psi^{(h)}(x, \theta, \bar \theta)\Big|_{\bar \theta = 0}, \nonumber\\
&& s_b \,s_{ab} \Psi(x) = \frac{\partial}{\partial \bar \theta}\,\frac{\partial}{\partial \theta}\, \tilde \Psi^{(h)}(x, \theta, \bar \theta).
\end{eqnarray}
It is worthwhile to mention that the (anti-)BRST transformations of the fermionic auxiliary fields $\rho$, $\lambda$ and 
Nakanishi--Lautrup type fields $B_\mu$, $\bar B_\mu$ have been derived from the requirements of the nilpotency and absolute
anticommutativity properties  of the (anti-)BRST  symmetry transformations.

So far, we have obtained the off-shell nilpotent and absolutely anticommuting (anti-) BRST symmetries for the 2-form  field $B_{\mu\nu}$ and
corresponding (anti-)ghost fields. But, the (anti-)BRST symmetry transformations of the St{\"u}ckelberg vector field $\phi_\mu$ 
and corresponding (anti-)ghost fields
are still unknown. For this purpose, it is to be noted that following quantity remains invariant under the gauge transformations 
$(\delta = \delta_1 + \delta_2)$:
\begin{eqnarray}
&& \delta \left[B_{\mu\nu} - \frac{1}{m}\,\big(\partial_\mu \phi_\nu - \partial_\nu \phi_\nu\big)\right] = 0.
\end{eqnarray}
This is a physical quantity in the sense that it is gauge-invariant. Thus, it remains independent of the Grassmannian variables when we 
generalize it on the $(4, 2)$D supermanifold.   
This gauge-invariant quantity will serve our purpose in deriving the proper (anti-)BRST transformations 
for the St{\"u}ckelberg-like vector field $\phi_\mu$ and corresponding (anti)ghost fields $(\bar C)C$. 
In terms of the differential forms, we generalize this gauge-invariant restriction on the $(4, 2)$D supermanifold as
\begin{eqnarray}
&& B^{(2)} - \frac{1}{m}\,  d \, \phi^{(1)} =\tilde {\cal B}^{(2)} - \frac{1}{m}\, \tilde d\, {\tilde {\bf \Phi}}^{(1)}, 
\end{eqnarray}
where the super 1-form is defined as
\begin{eqnarray}
{\tilde {\bf \Phi}}^{(1)} &=& dZ^M {\bf \Phi}_M \nonumber\\
&=& dx^\mu {\tilde {\bf  \Phi}}_\mu (x, \theta, \bar \theta) 
+ d \theta {\tilde {\bar {\cal F}}}(x, \theta, \bar \theta) + d\bar \theta {\tilde {\cal F}} (x, \theta, \bar \theta). 
\end{eqnarray} 
The multiplets of super 1-from, one can express, along the Grassmannian directions as  
\begin{eqnarray}
{\tilde {\bf \Phi}}_\mu (x, \theta, \bar\theta) &=& \phi_\mu(x) + \theta {\bar R}_\mu (x) +\bar\theta  R_\mu (x) 
+ i \theta\bar\theta  S_\mu (x),\nonumber\\
{\tilde {\cal F}}(x, \theta, \bar\theta) &=& C(x) + \theta {\bar B}_1 (x)  + \bar\theta  B_1 (x)  
+ i \theta\bar\theta\, s(x), \nonumber\\
{\tilde {\bar {\cal F}}} (x, \theta, \bar\theta) &=& \bar C (x) + \theta\,{\bar B}_2 (x)  +  \bar\theta  B_2 (x)  
+ i \theta\bar\theta \bar s(x),
\end{eqnarray}  
where $R_\mu, \bar R_\mu, s, \bar s$ and $S_\mu,  B_1, \bar B_1, B_2, \bar B_2$ are fermionic and bosonic secondary fields, respectively.

The explicit expression of the r.h.s. of (44) can be written in the following fashion: 
\begin{eqnarray} 
\tilde {\cal B}^{(2)} - \frac{1}{m} \, \tilde d \, \tilde{\bf  \Phi}^{(1)} &=& \frac{1}{2!} \big(dx^\mu \wedge dx^\nu\big) 
\Big[\tilde {\cal B}^{(h)}_{\mu\nu} - \frac{1}{m} \big(\partial_\mu \tilde {\bf \Phi}_\nu - \partial_\nu \tilde {\bf \Phi}_\nu\big)\Big]\nonumber\\
&+&  \big(dx^\mu \wedge d \theta\big) \Big[{\tilde{\bar {\cal F}}}_\mu^{(h)} 
- \frac{1}{m} \big(\partial_\mu {\tilde {\bar {\cal F}}} - \partial_\theta {\tilde {\bf \Phi}}_\mu \big) \Big] \nonumber\\
&+& \big(dx^\mu \wedge d \bar \theta\big) \Big[{\tilde {\cal F}}_\mu^{(h)} 
- \frac{1}{m}\big(\partial_\mu {\tilde {\cal F}} - \partial_{\bar\theta} \tilde {\bf \Phi}_\mu \big) \Big]\nonumber\\
&+& \big(d \theta \wedge d \bar \theta\big) \Big[{\tilde \Phi}^{(h)} 
+ \frac{1}{m} \big(\partial_\theta{\tilde  {\cal F}} + \partial_{\bar \theta} {\tilde {\bar {\cal F}}} \big)\Big] \nonumber\\
&+& \big(d \theta \wedge d \theta\big) \Big[{\tilde {\bar \beta}}^{(h)} + \frac{1}{m}\, \partial_\theta {\tilde {\bar {\cal F}}}\Big] 
+ \big(d \bar \theta \wedge d \bar \theta\big) \Big[{\tilde \beta}^{(h)}
+ \frac{1}{m}\, \partial_{\bar \theta}{\tilde  {\cal F}}\Big].
\end{eqnarray}
Using (44) and setting all the coefficients of the Grassmannian differentials to zero, we obtain the following interesting relationships: 
\begin{eqnarray} 
&& {\tilde{\bar {\cal F}}}_\mu^{(h)} - \frac{1}{m}\, \big(\partial_\mu {\tilde {\bar {\cal F}}} 
- \partial_\theta {\tilde {\bf \Phi}}_\mu \big) = 0, 
\qquad {\tilde {\cal F}}_\mu^{(h)} - \frac{1}{m}\, \big(\partial_\mu {\tilde {\cal F}} 
- \partial_{\bar\theta} \tilde {\bf \Phi}_\mu \big) = 0, \nonumber\\
&& {\tilde {\bar \beta}}^{(h)} + \frac{1}{m}\, \partial_\theta {\tilde {\bar {\cal F}}} = 0,\qquad
{\tilde \beta}^{(h)}  + \frac{1}{m}\, \partial_{\bar \theta}{\tilde  {\cal F}} = 0, \nonumber\\
&& {\tilde \Phi}^{(h)} + \frac{1}{m}\,\big(\partial_\theta{\tilde  {\cal F}} + \partial_{\bar \theta} {\tilde {\bar {\cal F}}} \big) = 0.
\end{eqnarray}
Exploiting the above equations together with (41) for the super-expansions given in (46), we obtain the precise value of the 
secondary fields in terms of the basic and auxiliary fields, namely;
\begin{eqnarray}
&& R_\mu = \partial_\mu C - m \,C_\mu, \quad \bar R_\mu = \partial_\mu \bar C - m\,\bar C_\mu, 
\quad B_1 = - m\, \beta, \quad \bar B_2 = - m\, \beta, \nonumber\\
&& B_2 + \bar B_1 + m \varphi = 0,\qquad s = i\,m\, \lambda, \qquad \bar s = - i\,m\, \rho, \nonumber\\
&& S_\mu = - i \,\big(\partial_\mu B_2 -  m \, B_\mu \big) \equiv + i\, \big(\partial_\mu \bar B_1 - m\, \bar B_\mu \big).
\end{eqnarray}
Putting the above relationships into the super-expansions of the superfields (46), we obtain the following explicit expressions 
for the superfields (46), in terms of the basic and auxiliary fields:
\begin{eqnarray}
{\tilde {\bf \Phi}}^{(h,g)}_\mu (x, \theta, \bar\theta) &=& \phi_\mu(x) + \theta \big(\partial_\mu \bar C - m\, \bar C_\mu\big)(x) 
+ \bar \theta \big(\partial_\mu C - m\, C_\mu\big)(x)\nonumber\\
&+& \theta\,\bar\theta \big(\partial_\mu B - m\, B_\mu\big) (x),\nonumber\\
{\tilde {\cal F}}^{(h,g)}(x, \theta, \bar\theta) &=& C(x) + \theta\,{\bar B} (x)  - \bar\theta \big(m\, \beta \big)(x)  
- \theta \, \bar\theta \big(m\, \lambda \big)(x), \nonumber\\
{\tilde {\bar {\cal F}}}^{(h,g)} (x, \theta, \bar\theta) &=& \bar C(x) - \theta \big(m \bar \beta \big) (x) +  \bar\theta B(x)  
+ \theta \, \bar\theta \big(m\, \rho \big)(x).
\end{eqnarray}  
Here the superscript $(h, g)$ on the superfields denotes the super-expansions obtained after the application of gauge-invariant restriction (44). 
In the above, we have made the choices $B_2 = B$ and $\bar B_1  = \bar B$ for the additional Nakanishi--Lautrup type fields $B$ and $\bar B$.
These fields are also required for the off-shell nilpotency of the (anti-) BRST symmetry transformations and linearization of the gauge-fixing term for the 
St{\"u}ckelberg vector field $\phi_\mu$. Again, the (anti-)BRST transformations for the auxiliary fields $B$ and $\bar B$ have been derived from the 
requirements of the nilpotency and absolute anticommutativity of the (anti-)BRST transformations. Thus, one can easily  read-off all the 
(anti-)BRST transformations for the vector field $\phi_\mu$ and corresponding (anti)ghost fields $(\bar C)C$ [cf. (10) and (11)].

Before we wrap up this section, we point out that the CF conditions (13) which play the crucial role (cf. Sect. 3) emerge very naturally 
in this formalism. The {\it first} CF condition $B_\mu + \bar B_\mu + \partial_\mu \varphi = 0$ arises from the HC (28). In particular, the  relation 
$\partial_\mu {\tilde { \Phi}} + \partial_\theta {\tilde {{\cal F}}}_\mu +  \partial_{\bar\theta} {\tilde {\bar {\cal F}}}_\mu = 0 $, which is 
a coefficient of  the wedge product $\big(dx^\mu \wedge d \theta \wedge d \bar \theta \big)$, leads to the first CF condition. Similarly,
 the {\it second} CF condition 
$B + \bar B  + m\, \varphi = 0$ emerges from (48) when we set the coefficient of the wedge product $(d \theta \wedge d \bar \theta)$ equal to zero. 
In fact, the last relation of Eq. (48) produces the second CF condition.
Furthermore, it is interesting to note that the equation (44) imposes its own integrability condition \cite{mb}. Thus, if we operate super exterior derivative 
$\tilde d = d + d\theta\, \partial_\theta + d \bar \theta\, \partial_{\bar \theta}$ on  (44) from left, we obtain
\begin{eqnarray}
\tilde d\, \Big( B^{(2)} - \,\frac{1}{m}\, d \, \phi^{(1)} \Big) 
= \tilde d \Big(\tilde {\cal B}^{(2)} - \frac{1}{m}\, \tilde d\, {\tilde {\bf \Phi}}^{(1)} \Big). 
\end{eqnarray}
In the above, $B^{(2)}$ and $\phi^{(1)}$ are independent of the Grassmannian variables $(\theta, \bar \theta)$ 
and $d^2 = \tilde d^2 = 0$. As a result, the above equation turns into the horizontality condition (37).


\section{(Anti-)BRST invariance of the coupled Lagrangian densities: superfield approach}

In this section, we shall provide the (anti-)BRST invariance of the coupled Lagrangian densities in the context of superfield formalism.
To accomplish this goal, we note that the coupled Lagrangian densities, in terms of the off-shell nilpotent and absolutely anticommuting (anti-)BRST
symmetries, can be written as
\begin{eqnarray}
{\cal L}_B =  {\cal L}_s  &+& s_b \, s_{ab} \bigg[\frac{1}{2}\, \phi_\mu \,\phi^\mu - \frac{1}{4}\, B_{\mu\nu}\, B^{\mu\nu} 
+ \bar C_\mu \, C^\mu + \frac{1}{2}\, \varphi \, \varphi + 2\bar \beta \,  \beta + C \,\bar C \bigg],
\end{eqnarray}
\begin{eqnarray}
{\cal L}_{\bar B} =  {\cal L}_s &-& s_{ab} \, s_b \bigg[\frac{1}{2}\, \phi_\mu \,\phi^\mu - \frac{1}{4}\, B_{\mu\nu}\, B^{\mu\nu} 
+ \bar C_\mu \, C^\mu + \frac{1}{2}\, \varphi \, \varphi + 2\bar \beta \,  \beta + C \,\bar C \bigg].
\end{eqnarray}
For our present 4D model, all terms in square brackets are chosen in such a way that  
each term carries mass dimension equal to $[M]^2$ in natural units ($\hbar = c = 1)$. In fact, the dynamical fields 
$B_{\mu\nu}, \phi_\mu, C_\mu, \bar C_\mu, \beta, \bar \beta, \varphi, C, \bar C$ have mass dimension equal to $[M]$.
The operation of $s_b$ and $s_{ab}$  on any generic field increases the mass dimension by one unit. In other words, 
$s_b$ and $s_{ab}$ carry mass dimension one. Furthermore, $s_b$ increases the ghost number by one unit when it operates 
on any generic field whereas $s_{ab}$ decreases the ghost number by one unit when it acts on any field. As a consequence, 
the above coupled Lagrangian densities are consistent with mass dimension and ghost number considerations. The constant 
numerals in front of each term are chosen for our algebraic convenience.  In full blaze of glory, the above Lagrangian densities 
(52) and (53) lead to (8) and (9), respectively, modulo the total spacetime derivatives.

In our earlier section (cf. Sect. 2), we have already mentioned that ${\cal L}_s$ is gauge-invariant 
and, thus, it remains invariant under the (anti-)BRST symmetries. As a consequence, both 
${\cal L}_B$ and ${\cal L}_{\bar B}$ given in (52) and (53) remain invariant under the operation of $s_{(a)b}$ due to  the nilpotency property
(i.e. $s_b^2 = 0,\; s^2_{ab} = 0$) of $s_{(a)b}$. In terms of the superfields (41), (50) and Grassmannian translational generators, 
we can generalize the 4D Lagrangian densities to super Lagrangian densities on the $(4, 2)$D supermanifold as   
\begin{eqnarray}
\tilde {\cal L}_B &=&  {\tilde {\cal L}}_s + \frac{\partial}{\partial \bar \theta} \, \frac{\partial}{\partial  \theta} 
\bigg[\frac{1}{2}\, {\tilde {\bf \Phi}}^{(h,g)}_\mu  \,{\tilde {\bf \Phi}}^{\mu {(h,g)}} 
- \frac{1}{4}\, \tilde {\cal B}^{(h)}_{\mu\nu}\, \tilde {\cal B}^{\mu\nu {(h)}} \nonumber\\
&& \hskip 1cm +\; {\tilde {\bar {\cal F}}}^{(h)}_\mu \, {\tilde {\cal F}}^{\mu {(h)}} 
+ \frac{1}{2}\, \tilde \Phi^{(h)} \, \tilde \Phi^{(h)} + 2 \tilde {\bar \beta}^{(h)} \,  \tilde \beta^{(h)} 
+ {\tilde {\cal F}}^{(h,g)} \, {\tilde {\bar {\cal F}}}^{(h,g)} \bigg],
\end{eqnarray}
\begin{eqnarray}
\tilde {\cal L}_{\bar B} &=&  {\tilde {\cal L}}_s - \frac{\partial}{\partial  \theta}\, \frac{\partial}{\partial \bar \theta}  
\bigg[\frac{1}{2}\, {\tilde {\bf \Phi}}^{(h,g)}_\mu \, {\tilde {\bf \Phi}}^{\mu {(h,g)}} 
- \frac{1}{4}\, \tilde {\cal B}^{(h)}_{\mu\nu}\, \tilde {\cal B}^{\mu\nu {(h)}} \nonumber\\
&& \hskip 1cm +\;{\tilde {\bar {\cal F}}}^{(h)}_\mu \, {\tilde {\cal F}}^{\mu {(h)}} 
+ \frac{1}{2}\, \tilde \Phi^{(h)} \, \tilde \Phi^{(h)} + 2 \tilde {\bar \beta}^{(h)} \,  \tilde \beta^{(h)} 
+ {\tilde {\cal F}}^{(h,g)} \, {\tilde {\bar {\cal F}}}^{(h,g)} \bigg],
\end{eqnarray}
where the super Lagrangian density $ \tilde {\cal L}_s$ is the generalization of the gauge-invariant Lagrangian density ${\cal L}_s$ on 
the $(4, 2)$D superspace. The former Lagrangian density is given as follows:    
\begin{eqnarray}
\tilde {\cal L}_s &=& \frac{1}{12} \, \tilde {\cal H}^{(h)}_{\mu\nu\eta}\, \tilde {\cal H}^{\mu\nu\eta{(h)}} 
- \frac{m^2}{4} \,\tilde {\cal B}^{(h)}_{\mu\nu}\, \tilde {\cal B}^{\mu\nu (h)}  
- \frac{1}{4}\, \tilde {\bf \Phi}^{\mu\nu(h,g)} \tilde {\bf \Phi}^{(h,g)}_{\mu\nu} 
+\frac{m}{2}\, \tilde {\cal B}^{(h)}_{\mu\nu} \, \tilde{\bf \Phi}^{\mu\nu(h,g)}. 
\end{eqnarray}
A straightforward computation shows that $\tilde {\cal L}_s$ is independent of the  Grassmannian variables $\theta$ and $\bar \theta$ 
(i.e. $\tilde {\cal L}_s = {\cal L}_s$) which shows that  ${\cal L}_s$ is gauge-invariant as well as (anti-)BRST invariant Lagrangian density. 
Mathematically, latter can be expressed in terms of the translational generators as   
\begin{eqnarray}
&& \frac{\partial}{\partial \bar \theta} \, \tilde {\cal L}_s = 0 \Rightarrow s_b\, {\cal L}_s = 0,\nonumber\\
&& \frac{\partial}{\partial \theta} \, \tilde {\cal L}_s = 0 \Rightarrow s_{ab}\, {\cal L}_s = 0.
\end{eqnarray}
It is clear from (54) and (55) together with (57), the followings are true, namely;
\begin{eqnarray}
&& \frac{\partial}{\partial \bar \theta} \, \tilde {\cal L}_B = 0 \Rightarrow s_b\, {\cal L}_B = 0\nonumber\\
&& \frac{\partial}{\partial \theta} \, \tilde {\cal L}_{\bar B} = 0 \Rightarrow s_{ab}\, {\cal L}_{\bar B} = 0,
\end{eqnarray}
which clearly show the (anti-)BRST invariance of the coupled Lagrangian densities within the framework superfield formalism.
The above equation is true  due to the validity of the nilpotency (i.e. $\partial^2_\theta = 0, \; \partial^2_{\bar \theta} = 0$) 
of the translational generators $\partial_\theta$ and $\partial_{\bar\theta}$.


\section{Conclusions}

In our present investigation, we have studied the 4D gauge-invariant massive Abelian 2-form theory within the framework of BRST formalism where
the local gauge symmetries given in (7) are traded with two linearly independent global BRST and anti-BRST symmetries  [cf. (10) and (11)]. 
In this formalism, we have obtained the coupled (but equivalent) Lagrangian densities [cf. (8) and (9)] which respect the off-shell nilpotent and  
absolutely anticommuting BRST and anti-BRST symmetry transformations on the constrained hypersurface defined by the CF type conditions (13).     
These CF conditions  are (anti-)BRST invariant as well as they also play a pivotal role in the proof of  the absolute anticommutativity of the (anti-)BRST 
transformations and the derivation of the coupled Lagrangian densities. The anticommutativity property for the dynamical fields $B_{\mu\nu}$ and $\phi_\mu$ 
is satisfied due to the existence of the Curci--Ferrari type conditions [cf. (16)].

The continuous and off-shell nilpotent (anti-)BRST symmetries lead to the derivation of the corresponding 
conserved (anti-)BRST charges.  In addition to these symmetries, the coupled Lagrangian densities also respect 
the global ghost-scale symmetry which leads to the conserved ghost charge. The operator form of 
the continuous symmetry transformations and corresponding generators obeys the standard graded BRST algebra
[cf. (29) and (30)]. We lay emphasis on the fact that the physicality criteria $Q_{(a)b} |phys \rangle = 0$ 
produce the first-class constraints, as the physical conditions (24) on the theory, which are present in the 
gauge-invariant Lagrangian density (5). Thus, the BRST quantization is consistent with the Dirac quantization 
of the system having first-class constraints.

It is worthwhile to point out that the (anti-)BRST charges which are the generators of the corresponding symmetry transformations are 
unable to produce the proper (anti-) BRST symmetry transformations for the Nakanishi--Lautrup fields $B$, $\bar B$, $B_\mu$, $\bar B_\mu$ 
and other fermionic auxiliary fields $\rho$, $\lambda$. The transformations of these fields have been derived from the requirements of the 
nilpotency and absolute anticommutativity properties of the (anti-) BRST transformations. Similarly, the ghost charge is also incapable to generate 
the proper transformations for the auxiliary ghost fields $\rho$ and $\lambda$.  We have derived these symmetries from other considerations 
[cf. (30)] where we have used the appropriate relations that appear in the algebra (29).

Furthermore, we have exploited the augmented version of superfield approach to BRST formalism to derive the off-shell nilpotent and absolutely anticommuting 
(anti-)BRST symmetries for the 4D dimensional St{\"u}ckelberg-like massive Abelian 2-form gauge theory. In this approach, besides 
the horizontality condition (37), we have used the  gauge-invariant restriction (44) for the derivation of the complete sets of the BRST and 
anti-BRST symmetry transformations. The gauge-invariant restriction is required for the derivation of the proper (anti-)BRST transformations 
for the St{\"u}ckelberg-like vector field $\phi_\mu$. One of the spectacular observations, we point out that the horizontality condition,  
which produces the (anti-)BRST transformations for the 2-form field and corresponding (anti-)ghost fields, can also be obtained 
from the integrability of (44) \cite{mb}. The CF conditions, which are required for the absolute anticommutativity of the (anti-)BRST 
symmetries, emerge very naturally in the superfield formalism. These (anti-)BRST invariant CF conditions are conserved quantities. 
Thus, it would be an interesting piece of work to show that these CF conditions commute with the 
Hamiltonian within the framework of BRST formalism (see, e.g. \cite{mms,sbm}).

Using the basic tenets of BRST formalism, we have written the coupled (but equivalent) Lagrangian densities 
in terms of (anti-)BRST symmetries where the mass dimension and ghost number of the dynamical fields have been taken into account.
Within the framework of superfield, we have provided the geometrical origin
of the (anti-)BRST symmetries in terms of the Grassmannian translational generators. Also, one can capture the basic properties of the (anti-)BRST transformations in the language of the translational generators.  Thus, we have been able to write the coupled 
Lagrangian densities in terms of the superfields and Grassmannian derivatives. 
As a result, the (anti-)BRST invariance of the super Lagrangian densities become quite simpler and straightforward 
due to the nilpotency property of  the Grassmannian derivatives.

\section*{Acknowledgments} RK would like to thank UGC, Government of India, New Delhi, for financial support under the PDFSS scheme.
SK acknowledges DST research grant EMR/2014/000250 for post-doctoral support.

\section*{Appendix A: Anticommutativity of the BRST and anti-BRST charges}
In this appendix, we provide an explicit proof of the anticommutativity of the  BRST and anti-BRST charges in 
a simpler way. The BRST and anti-BRST charges given in (20) can also be simplified by using the 
equations of motion (18) and (19), respectively, as
\begin{eqnarray}
Q_b &=& \int d^3x \bigg[B_i \big(\partial^0 C^i - \partial^i C^0 \big) - \big(\partial^0 B^i - \partial^i B^0 \big) C_i 
- B \big(\partial^0 C - m C^0 \big) \nonumber\\
&+& \big(\partial^0 B - m B^0 \big) C - \rho\big(\partial^0 \beta\big) + \beta \big(\partial^0\rho \big) + \lambda B^0  \bigg],
\end{eqnarray}
\begin{eqnarray}
Q_{ab} &=& \int d^3x \bigg[- \bar B_i \big(\partial^0 \bar C^i - \partial^i \bar C^0 \big) 
+ \big(\partial^0 \bar B^i - \partial^i \bar B^0 \big) \bar C_i 
+ \bar B \big(\partial^0 \bar C - m \bar C^0 \big) \nonumber\\
&-& \big(\partial^0 \bar B - m \bar B^0 \big)\bar  C 
- \lambda \big(\partial^0 \bar \beta\big) + \bar \beta \big(\partial^0 \lambda \big) + \rho \bar B^0  \bigg].
\end{eqnarray}
Applying $s_{ab}$ on $Q_b$ and using the equation of motion for the ghost field $C_0$ [cf. (18)], we obtain
\begin{eqnarray}
s_{ab}\, Q_b&=& \int d^3x \bigg[B_i \big(\partial^0 \bar B^i - \partial^i \bar B^0 \big)
- \bar B_i \big(\partial^0 B^i - \partial^i  B^0 \big) \nonumber\\
&-& B \big(\partial^0 \bar B - m \bar B^0 \big) + \bar B \big(\partial^0 B - m B^0 \big)  \bigg].
\end{eqnarray}
Now eliminating $\bar B^i$, $\bar B^0$ and $\bar B$ by using the CF conditions,  the above expression further simplifies as
\begin{eqnarray}
s_{ab}\, Q_b &=& \int d^3x \bigg[\varphi\, \big(\partial_i \partial^i B^0 + m^2\, B^0 \big)
- \varphi\, \partial^0\big(\partial_i B^i + m\, B \big) \bigg].
\end{eqnarray}
Exploiting the equation of motion  $\partial_\mu B^\mu + m \, B = 0$ and an off-shoot $(\Box + m^2)\,B_\mu =0$ of the Euler--Lagrange  equations 
of motion (18), we obtain
\begin{eqnarray}
s_{ab}\, Q_b = -i \big\{Q_b, \, Q_{ab}\} = 0.
\end{eqnarray} 
Similarly, operating BRST transformations $s_b$ on (A.2) and exploiting the equation of motion for the anti-ghost field $\bar C_0$ [cf. (18)], 
we finally obtain the r.h.s. of (A.3). In fact, we yield 
\begin{eqnarray}
s_b\,Q_{ab} = s_{ab}\, Q_{b}.
\end{eqnarray}
As a result of the above equations, the relation $s_b\,Q_{ab} = s_{ab}\, Q_b = - i \big\{Q_b, \, Q_{ab}\big\} = 0$ implies  
the anticommutativity (i.e. $Q_b\,Q_{ab} + Q_{ab}\, Q_b = 0$) of the (anti-)BRST charges $Q_{(a)b}$.
Here, we again lay emphasis on the fact that the CF conditions play a crucial 
role in the anticommutativity of the (anti-)BRST charges (and corresponding symmetry transformations).

\end{document}